\documentclass[a4paper,UKenglish,cleveref, autoref, thm-restate]{lipics-v2019}

\usepackage{float}
\usepackage{leftidx}
\usepackage{xcolor}
\usepackage{dsfont}
\usepackage{stmaryrd}
\usepackage{wasysym}
\usepackage{textgreek}
\usepackage{proof}
\usepackage{mathpartir}
\usepackage{sfmath}
\usepackage{url}
\usepackage{marvosym}
\usepackage{appendix}
\usepackage{pdfpages}
\usepackage{xifthen}
\usepackage{cleveref}
\crefformat{footnote}{#2\footnotemark[#1]#3}

\usepackage{tikz}
\usetikzlibrary{cd,decorations.markings,intersections}
\usetikzlibrary{arrows}
\tikzset{
    endarrow/.style={
        -,
        postaction={decorate,-},
        decoration={
            markings,
            mark=at position 1 with {
                \arrow{Triangle[fill=black]}
            }
        }
    }
}
\tikzset{
    midarrow/.style={
        -,
        postaction={decorate,-},
        decoration={
            markings,
            mark=at position .7 with {
                \arrow{Triangle[fill=black]}
            }
        }
    }
}

\definecolor{altlightcolour}{gray}{.6}
\definecolor{altmidcolour}{gray}{.5}
\definecolor{altdarkcolour}{gray}{.3}
\newcommand{\altlcm}[1]{\textcolor{altlightcolour}{\ensuremath{#1}}}
\newcommand{\altmcm}[1]{\textcolor{altmidcolour}{\ensuremath{#1}}}

\newcommand{\mathemph}[1]{\altmcm{\underline{\mathrm{#1}}}}

\newcommand{\N}{\mathbb{N}}
\newcommand{\inl}{\mathsf{inl}}
\newcommand{\inr}{\mathsf{inr}}
\newcommand{\finset}[1]{\underline{#1}}
\newcommand{\src}{\mathsf{src}}
\newcommand{\trg}{\mathsf{trg}}
\newcommand{\defeq}{:\equiv}
\newcommand{\bool}{\finset{2}}

\newcommand{\id}{\mathsf{id}}
\newcommand{\linj}{\mathsf{left\text{-}inj}}
\newcommand{\tolinj}{\lhook\joinrel\xrightarrow{\mathsf{left}}}
\newcommand{\dimens}{\mathsf{dim}}
\newcommand{\dimpres}{\mathsf{dim\text{-}pres}}
\newcommand{\diminj}{\mathsf{dim\text{-}inj}}
\newcommand{\isinj}{\mathsf{is\text{-}inj}}
\newcommand{\issurj}{\mathsf{is\text{-}surj}}
\newcommand{\obj}[1]{\mathsf{obj}(#1)}
\newcommand{\mor}[1]{#1}
\newcommand{\step}[1]{\text{[S#1]}}
\newcommand{\negate}{\mathsf{neg}}
\newcommand{\intv}{\ensuremath{\mathsf{intv}}}
\newcommand{\slimstar}{\!\star{}\!}

\newcommand{\graphhom}[2]{\mathsf{grp\text{-}hom}\left(#1,#2\right)}
\newcommand{\graphcube}{\square_{\mathsf{grp}}}
\newcommand{\graphmeet}{\square_{\mathsf{cont}}}
\newcommand{\graphdim}{\square_{\mathsf{dim}}}
\newcommand{\graphprism}{\mathsf{prism}}

\newcommand{\cube}[1]{C_{#1}}
\newcommand{\shcube}{\square_{\mathsf{BCH}}} 
\newcommand{\shcubeop}{\square^\mathrm{op}_{\mathsf{BCH}}} 
\newcommand{\meet}{\sqcap}
\newcommand{\join}{\sqcup}
\newcommand{\presmeet}{\mathsf{pres\text{-}meet}}
\newcommand{\presjoin}{\mathsf{pres\text{-}join}}
\newcommand{\twcubecat}{\Bowtie_\mathsf{grp}}
\newcommand{\twgraphprism}{\mathsf{tw\text{-}prism}}
\newcommand{\twcube}[1]{T_{#1}}
\newcommand{\twsquare}{\Bowtie}

\newcommand{\twgraphdim}{\twsquare_\mathsf{dim}}

\newcommand{\twternary}{\twsquare_\mathsf{tri}}

\title{From Cubes to Twisted Cubes \newline via Graph Morphisms in Type Theory}
\titlerunning{From Cubes to Twisted Cubes via Graph Morphisms in Type Theory}

\author{Gun Pinyo}{School of Computer Science, University of Nottingham, United Kingdom}{gunpinyo@gmail.com}{https://orcid.org/0000-0002-8483-5261}{}

\author{Nicolai Kraus}{School of Computer Science, University of Nottingham, United Kingdom \and \url{https://nicolaikraus.github.io/}}{nicolai.kraus@nottingham.ac.uk}{https://orcid.org/0000-0002-8729-4077}{The Royal Society, grant No.~URF\textbackslash R1\textbackslash 191055.}

\authorrunning{G. Pinyo and N. Kraus}

\Copyright{Gun Pinyo and Nicolai Kraus}



\relatedversion{This paper is also available at \url{https://arxiv.org/abs/1902.10820}.}

\EventEditors{Marc Bezem and Assia Mahboubi}
\EventNoEds{2}
\EventLongTitle{25th International Conference on Types for Proofs and Programs (TYPES 2019)}
\EventShortTitle{TYPES 2019}
\EventAcronym{TYPES}
\EventYear{2019}
\EventDate{June 11--14, 2019}
\EventLocation{Oslo, Norway}
\EventLogo{}
\SeriesVolume{175}
\ArticleNo{5}

\nolinenumbers{}

\begin{document}

\maketitle

\begin{abstract}
  \emph{Cube categories} are used to encode higher-dimensional
  categorical structures.  They have recently gained significant
  attention in the community of homotopy type theory and univalent
  foundations, where types carry the structure of higher groupoids.
  Bezem, Coquand, and Huber~\cite{bezem_et_al:LIPIcs:2014:4628} have
  presented a constructive model of univalence using a specific cube
  category, which we call the \emph{BCH cube category}.
  
  The higher categories encoded with the BCH cube category have the
  property that all morphisms are invertible, mirroring the fact that
  equality is symmetric.  This might not always be desirable: the
  field of \emph{directed type theory} considers a notion of equality
  that is not necessarily invertible.
  
  This motivates us to suggest a category of \emph{twisted cubes}
  which avoids built-in invertibility.  Our strategy is to first
  develop several alternative (but equivalent) presentations of the
  BCH cube category using morphisms between suitably defined graphs.
  Starting from there, a minor modification allows us to define our
  category of twisted cubes.  We prove several first results about
  this category, and our work suggests that twisted cubes combine
  properties of cubes with properties of globes and simplices
  (tetrahedra).
\end{abstract}

\section{Introduction and Motivation}

A \emph{cube category} is a category whose objects are (or represent)
finite-dimensional cubes, and whose morphisms are mappings of some
sort between these cubes.  There are many different cube
categories~\cite{oriended-cube, Antolini2002,
  bezem_et_al:LIPIcs:2014:4628,
  Ed-Morehouse-Varieties-of-Cubical-Sets,
  Clive-Newstead-cubical-sets}, and they are used to encode higher
categorical structures.

\emph{Homotopy type theory}~\cite{hottbook} is a variation of
Martin-L\"of's intensional type theory.  The characteristic and novel
view adapted in homotopy type theory is that types carry the structure
of higher categories, or, to be precise, higher groupoids (i.e.\ all
morphisms are invertible).  This view supports Voevodsky's
\emph{univalence principle} which should be seen as a central concept
of homotopy type theory.  The first model of such a type theory, given
by Voevodsky~\cite{voevodsky_univalentFoundationsProjectNSF} (see also
the presentation by Kapulkin and
Lumsdaine~\cite{kapulkin2012simplicial}), uses \emph{simplicial sets}.
However, it is still an open question how simplicial sets can be used
to build a \emph{constructive} model of type theory with univalent
universes~\cite{gambino2017frobenius}.  Using \emph{cubical sets},
this has been achieved by Bezem, Coquand, and
Huber~\cite{bezem_et_al:LIPIcs:2014:4628}.  Starting from there, cubes
have gathered a lot of attention in the type theory community, leading
to various \emph{cubical type theories} which have univalence not as
an axiom but as a built-in derivable
principle~\cite{CartesianCubicalTypeTheory, awodey2018cubical,
  cubicaltt, PittsAM:aximct-jv}.  Many different cube categories have
been considered in this context.

The important cube category used by Bezem, Coquand, and
Huber~\cite{bezem_et_al:LIPIcs:2014:4628} (from now on referred to as
the \emph{BCH cube category}) uses finite sets of variable names as
objects, and a morphism from a set $I$ to a set $J$ is a function
$f : I \to J \cup \{0,1\}$ which is ``injective on the left part'',
i.e.\ $f(i_1) = f(i_2) = j$ with $j : J$ implies $i_1 = i_2$.  One
goal of this paper is to develop several alternative presentations of
this category, mainly using graph morphisms.  We have two main
motivations to do this.  The first is that, as we hope, our
alternative and intuitive (but equivalent) definitions enable new
views on the category and facilitate the discovery of further
observations.  The second motivation is that a minor change in the
definition will allow us to construct a new cube category, the
\emph{twisted cubes} from the title. We will come back to this in a
moment.

\begin{figure}
\begin{center}
\begin{tikzpicture}[x=1.5cm,y=1.25cm,baseline=(current bounding box.center)]
\node (L) at (0,0) {$00$};
\node (M) at (0,1) {$01$};
\node (N) at (1,0) {$10$};
\node (P) at (1,1) {$11$};

\draw[->,endarrow] (L) to node {} (M);
\draw[->,endarrow] (L) to node {} (N);
\draw[->,endarrow, dashed] (M) to node {} (P);
\draw[->,endarrow] (N) to node {} (P);
\end{tikzpicture}
\hspace{1.5cm}
\begin{tikzpicture}[x=1.5cm,y=1.25cm,baseline=(current bounding box.center)]
\node (L) at (0,0) {$x$};
\node (M) at (0,1) {$y$};
\node (N) at (1,0) {$x$};
\node (P) at (1,1) {$x$};

\draw[->,endarrow] (L) to node [left] {$\scriptstyle p$} (M);
\draw[->,endarrow] (L) to node [above] {$\scriptstyle \id_x$} (N);
\draw[->,endarrow, dashed] (M) to node {} (P);
\draw[->,endarrow] (N) to node [right] {$\scriptstyle \id_x$} (P);
\end{tikzpicture}
\hspace{1.5cm}
\begin{tikzpicture}[x=1.5cm,y=1.25cm,baseline=(current bounding box.center)]
\node (L) at (0,0) {$00$};
\node (M) at (0,1) {$01$};
\node (N) at (1,0) {$10$};
\node (P) at (1,1) {$11$};

\draw[<-,endarrow] (M) to node {} (L);
\draw[->,endarrow] (L) to node {} (N);
\draw[->, dashed,endarrow] (M) to node {} (P);
\draw[->,endarrow] (N) to node {} (P);
\end{tikzpicture}

\caption{Kan-filling condition of a $2$-cube (left), a proof of
  invertibility introduced by the Kan-filling condition (middle), and
  how to remove such the invertibility (right).}~\label{fig:intro}
\end{center}
\end{figure}
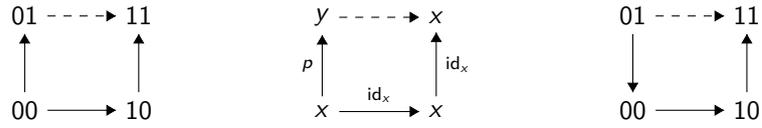

The standard way to create models (of both higher categories and type
theories) using simplicial or cubical index categories is to take
presheaves and equip them with certain \emph{Kan-filling conditions}.
These filling conditions entail composition of morphisms as well as
associativity and all higher coherence laws that one needs.  A typical
such Kan-filling condition for the $2$-cube%
\footnote{While Bezem, Coquand, and
  Huber~\cite{bezem_et_al:LIPIcs:2014:4628} define their index
  category to have finite sets of variables as objects, it is possible
  to simply use natural numbers as objects.  The \emph{$n$-cube}, or
  \emph{$n$-dimensional cube}, is then the object of the presheaf
  category that is represented by the object $n$ of the index
  category.}, as shown on the left of \cref{fig:intro}, says that,
given the ``partial square'' of three solid edges on the right, one
can always find the dashed edge (together with an actual filler for
the square).

One important observation here is that, in the case of the BCH cube
category and other cube categories, invertibility of morphisms is
built-in.  Consider the partial square, as shown on the middle of
\cref{fig:intro}, where two of the three solid edges are identities
and the third is an actual non-trivial morphism (or equality) $p$ from
$x$ to $y$.  Using the Kan filling operation described above, we get a
morphism from $y$ to $x$, which serves as the inverse of $p$.

The invertibility of morphisms is useful for most forms of type
theory, where equaliy is symmetric.  This however is not always the
case, cf.~the proposals for \emph{directed type theories} by Licata
and Harper~\cite{licata20112}, Nuyts~\cite{AndreasNuytsThesis}, Riehl
and Shulman~\cite{riehl2017type}, North~\cite{north2018towards}, and
others.  Their aim is to generalise type theory by replacing
\emph{(higher) groupoids} by general \emph{(higher) categories}.  In a
nutshell, this means that ``equality'' (or whatever takes the place of
equality) is not necessarily invertible.

We think that a very valuable long-term goal would be to make the
connection of directed type theories with cubical type theories and
create some sort of \emph{directed cubical type theory}.  This is at
the moment certainly out of reach, and we do not know how such a type
theory could be built.  Nevertheless, it motivates us to explore
variations of the BCH cube category which do not have the described
built-in equality.

To avoid invertibility, we ``twist'' the left-most edge of the
$2$-dimensional cube, as shown on the right of \cref{fig:intro}, to
ensure that the construction from before becomes impossible. This
might seems artificial and specific to the $2$-dimensional case but by
using our graph morphisms that we develop for the BCH cube category,
it becomes very easy to define the twisting version for cubes of all
dimensions.

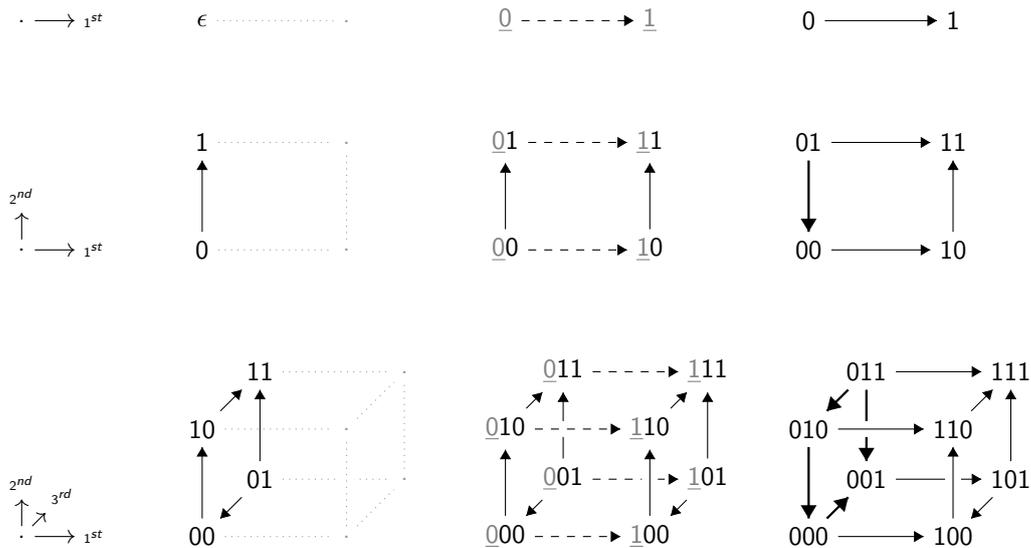
\begin{figure}
\begin{center}
\begin{tikzpicture}[x=.95cm,y=.95cm,baseline=(current bounding box.center)]
  \pgfmathsetmacro\xdif{2}    
  \pgfmathsetmacro\ydif{1.5}  
  \pgfmathsetmacro\zxdif{0.8} 
  \pgfmathsetmacro\zydif{0.8} 
  \pgfmathsetmacro\rxy{0.5}   
  \pgfmathsetmacro\rz{0.7}    
  \pgfmathsetmacro\xbd{4.2}   
  \pgfmathsetmacro\ybd{-4}    
  \pgfmathsetmacro\yfstd{-.8}  
  \foreach \d in {0, 1, 2}{   
    \pgfmathsetmacro\xb{-2.5}
    \pgfmathsetmacro\yb{\ybd*\d}
    \ifthenelse{\cnttest{\d}={0}}{
      \pgfmathsetmacro\yb{\ybd*\d + \yfstd}
    }{}
    \node (D\d)  at (\xb,              \yb             ) {$.$};
    \node (DX\d) at (\xb + \rxy*\xdif, \yb             ) {\tiny{$1^{st}$}};
    \draw[->] (D\d) to node {} (DX\d);
    \ifthenelse{\cnttest{\d}>{0}}{
      \node (DY\d) at (\xb             , \yb + \rxy*\ydif) {\tiny{$2^{nd}$}};
      \draw[->] (D\d) to node {} (DY\d);
    }{}
    \ifthenelse{\cnttest{\d}>{1}}{
      \node (DZ\d) at (\xb + \rz*\zxdif, \yb + \rz*\zydif) {\tiny{$3^{rd}$}};
      \draw[->] (D\d) to node {} (DZ\d);
    }{}
  }

  \pgfmathsetmacro\yb{0*\ybd + \yfstd}
  \pgfmathsetmacro\xb{0*\xbd}
  \node (N00P0) at (\xb        , \yb) {\ensuremath{\epsilon}};
  \node (N00P1) at (\xb + \xdif, \yb) {\altlcm{.}};
  \draw[altlightcolour, dotted] (N00P0) to node {} (N00P1);
  \pgfmathsetmacro\xb{\xbd*1}
  \node (N01P0) at (\xb        , \yb) {\mathemph{0}};
  \node (N01P1) at (\xb + \xdif, \yb) {\mathemph{1}};
  \draw[->, endarrow, dashed]           (N01P0) to node {} (N01P1);
  \pgfmathsetmacro\xb{\xbd*2}
  \node (N02P0) at (\xb        , \yb) {\ensuremath{0}};
  \node (N02P1) at (\xb + \xdif, \yb) {\ensuremath{1}};
  \draw[->, endarrow]           (N02P0) to node {} (N02P1);
  
  \pgfmathsetmacro\yb{\ybd*1}
  \pgfmathsetmacro\xb{\xbd*0}
  \node (N10P00) at (\xb        , \yb        ) {\ensuremath{0}};
  \node (N10P01) at (\xb        , \yb + \ydif) {\ensuremath{1}};
  \node (N10P10) at (\xb + \xdif, \yb        ) {\altlcm{.}};
  \node (N10P11) at (\xb + \xdif, \yb + \ydif) {\altlcm{.}};
  \draw[altlightcolour, dotted] (N10P00) to node {} (N10P10);
  \draw[altlightcolour, dotted] (N10P01) to node {} (N10P11);
  \draw[->, endarrow]           (N10P00) to node {} (N10P01);
  \draw[altlightcolour, dotted] (N10P10) to node {} (N10P11);
  
  \pgfmathsetmacro\xb{\xbd*1}
  \node (N11P00) at (\xb        , \yb        ) {\ensuremath{\mathemph{0}0}};
  \node (N11P01) at (\xb        , \yb + \ydif) {\ensuremath{\mathemph{0}1}};
  \node (N11P10) at (\xb + \xdif, \yb        ) {\ensuremath{\mathemph{1}0}};
  \node (N11P11) at (\xb + \xdif, \yb + \ydif) {\ensuremath{\mathemph{1}1}};
  \draw[->, endarrow, dashed]                     (N11P00) to node {} (N11P10);
  \draw[->, endarrow, dashed]                     (N11P01) to node {} (N11P11);
  \draw[->, endarrow]                     (N11P00) to node {} (N11P01);
  \draw[->, endarrow]                     (N11P10) to node {} (N11P11);
  
  \pgfmathsetmacro\xb{2*\xbd}
  \node (N12P00) at (\xb        , \yb        ) {\ensuremath{00}};
  \node (N12P01) at (\xb        , \yb + \ydif) {\ensuremath{01}};
  \node (N12P10) at (\xb + \xdif, \yb        ) {\ensuremath{10}};
  \node (N12P11) at (\xb + \xdif, \yb + \ydif) {\ensuremath{11}};
  \draw[->, endarrow]                     (N12P00) to node {} (N12P10);
  \draw[->, endarrow]                     (N12P01) to node {} (N12P11);
  \draw[line width=0.3mm, endarrow]       (N12P01) to node {} (N12P00);
  \draw[->, endarrow]                     (N12P10) to node {} (N12P11);
    
  \pgfmathsetmacro\yb{\ybd*2}
  \pgfmathsetmacro\xb{\xbd*0}
  \node (N20P000) at (\xb                 , \yb                 )
      {\ensuremath{00}};
  \node (N20P001) at (\xb +         \zxdif, \yb +         \zydif)
      {\ensuremath{01}};
  \node (N20P010) at (\xb                 , \yb + \ydif         )
      {\ensuremath{10}};
  \node (N20P011) at (\xb +         \zxdif, \yb + \ydif + \zydif)
      {\ensuremath{11}};
  \node (N20P100) at (\xb + \xdif         , \yb                 )
      {\altlcm{.}};
  \node (N20P101) at (\xb + \xdif + \zxdif, \yb +         \zydif)
      {\altlcm{.}};
  \node (N20P110) at (\xb + \xdif         , \yb + \ydif         )
      {\altlcm{.}};
  \node (N20P111) at (\xb + \xdif + \zxdif, \yb + \ydif + \zydif)
      {\altlcm{.}};
  \draw[<-,endarrow] (N20P000) to node {} (N20P010);
  \draw[<-,endarrow] (N20P001) to node {} (N20P011);
  \draw[->,endarrow] (N20P001) to node {} (N20P000);
  \draw[<-,endarrow] (N20P010) to node {} (N20P011);
  \draw[altlightcolour, dotted] (N20P000) to node {} (N20P100);
  \draw[altlightcolour, dotted] (N20P011) to node {} (N20P111);
  \draw[altlightcolour, dotted] (N20P101) to node {} (N20P111);
  \draw[altlightcolour, dotted] (N20P100) to node {} (N20P101);
  \draw[altlightcolour, dotted] (N20P110) to node {} (N20P111);
  \draw[altlightcolour, dotted] (N20P001) to node {} (N20P101);
  \draw[altlightcolour, dotted] (N20P010) to node {} (N20P110);
  \draw[-,color=white,line width=6pt] (N20P100) to node {} (N20P110);
  \draw[altlightcolour, dotted] (N20P100) to node {} (N20P110);
  \pgfmathsetmacro\yb{\ybd*2}
  \pgfmathsetmacro\xb{\xbd*1}
  \node (N21P000) at (\xb                 , \yb                 )
      {\ensuremath{\mathemph{0}00}};
  \node (N21P001) at (\xb +         \zxdif, \yb +         \zydif)
      {\ensuremath{\mathemph{0}01}};
  \node (N21P010) at (\xb                 , \yb + \ydif         )
      {\ensuremath{\mathemph{0}10}};
  \node (N21P011) at (\xb +         \zxdif, \yb + \ydif + \zydif)
      {\ensuremath{\mathemph{0}11}};
  \node (N21P100) at (\xb + \xdif         , \yb                 )
      {\ensuremath{\mathemph{1}00}};
  \node (N21P101) at (\xb + \xdif + \zxdif, \yb +         \zydif)
      {\ensuremath{\mathemph{1}01}};
  \node (N21P110) at (\xb + \xdif         , \yb + \ydif         )
      {\ensuremath{\mathemph{1}10}};
  \node (N21P111) at (\xb + \xdif + \zxdif, \yb + \ydif + \zydif)
      {\ensuremath{\mathemph{1}11}};
  \draw[->,endarrow, dashed] (N21P000) to node {} (N21P100);
  \draw[->,endarrow, dashed] (N21P011) to node {} (N21P111);
  \draw[<-,endarrow] (N21P000) to node {} (N21P010);
  \draw[<-,endarrow] (N21P001) to node {} (N21P011);
  \draw[->,endarrow] (N21P101) to node {} (N21P111);
  \draw[->,endarrow] (N21P001) to node {} (N21P000);
  \draw[<-,endarrow] (N21P010) to node {} (N21P011);
  \draw[<-,endarrow] (N21P101) to node {} (N21P100);
  \draw[->,endarrow] (N21P110) to node {} (N21P111);
  \draw[->,endarrow, dashed] (N21P001) to node {} (N21P101);
  \draw[-,color=white,line width=6pt] (N21P010) to node {} (N21P110);
  \draw[->,endarrow, dashed] (N21P010) to node {} (N21P110);
  \draw[-,color=white,line width=6pt] (N21P100) to node {} (N21P110);
  \draw[->,endarrow] (N21P100) to node {} (N21P110);

  \pgfmathsetmacro\xb{\xbd*2}
  \node (N22P000) at (\xb                 , \yb                 )
      {\ensuremath{000}};
  \node (N22P001) at (\xb +         \zxdif, \yb +         \zydif)
      {\ensuremath{001}};
  \node (N22P010) at (\xb                 , \yb + \ydif         )
      {\ensuremath{010}};
  \node (N22P011) at (\xb +         \zxdif, \yb + \ydif + \zydif)
      {\ensuremath{011}};
  \node (N22P100) at (\xb + \xdif         , \yb                 )
      {\ensuremath{100}};
  \node (N22P101) at (\xb + \xdif + \zxdif, \yb +         \zydif)
      {\ensuremath{101}};
  \node (N22P110) at (\xb + \xdif         , \yb + \ydif         )
      {\ensuremath{110}};
  \node (N22P111) at (\xb + \xdif + \zxdif, \yb + \ydif + \zydif)
  {\ensuremath{111}};
  \draw[->,endarrow] (N22P000) to node {} (N22P100);
  \draw[->,endarrow] (N22P011) to node {} (N22P111);
  \draw[line width=0.3mm,<-,endarrow] (N22P010) to node {} (N22P000);
  \draw[line width=0.3mm,<-,endarrow] (N22P011) to node {} (N22P001);
  \draw[->,endarrow] (N22P101) to node {} (N22P111);
  \draw[line width=0.3mm,->,endarrow] (N22P000) to node {} (N22P001);
  \draw[line width=0.3mm,<-,endarrow] (N22P011) to node {} (N22P010);
  \draw[<-,endarrow] (N22P101) to node {} (N22P100);
  \draw[->,endarrow] (N22P110) to node {} (N22P111);
  \draw[->,endarrow] (N22P001) to node {} (N22P101);
  \draw[-,color=white,line width=6pt] (N22P010) to node {} (N22P110);
  \draw[->,endarrow] (N22P010) to node {} (N22P110);
  \draw[-,color=white,line width=6pt] (N22P100) to node {} (N22P110);
  \draw[->,endarrow] (N22P100) to node {} (N22P110);

\end{tikzpicture}
\caption{An illustration of the \emph{thickening-and-twisting} process
  of the twisted $n$-cube for $1 \leqslant n \leqslant 3$. The process
  expands the twisted $(n-1)$-cube (left column) along the new
  dimension (middle column) and reverse all other dimensions at the
  starting point of the new dimension (right
  column).}~\label{fig:thickening-and-twisting-process}
\end{center}
\end{figure}

To construct a twisted $n$-cube from a twisted $(n-1)$-cube, we first
expand the original cube along a new dimension (we call this
\emph{thickening}): this is same as constructing a standard $n$-cube
form a standard $(n-1)$-cube, which is just a construction of its
\emph{cylinder object}. We then reverse all dimensions at the starting
point of the new dimension (we call this \emph{twisting}).
\Cref{fig:thickening-and-twisting-process} illustrates this
\emph{thickening-and-twisting} process up to dimension $3$, where the
existing dimensions are shifted by one in order to allow the new
dimension to be the first dimension.

One important property of standard cubes which twisted cubes retain is
that every face of a [twisted] $n$-cube is a [twisted] $(n-1)$-cube.
An interesting example is the case $n = 3$: In order to construct a
twisted $3$-cube, we thicken the twisted $2$-cube as illustrate in
\cref{fig:thickening-and-twisting-process} where the left and the
right face are already twisted $2$-cubes, while the rest are thickened
$1$-cubes.  The right face is unaffected during the twisting, but the
left face is reversed entirely.  Nevertheless, it is still a $2$-cube
(just flipped backwards).

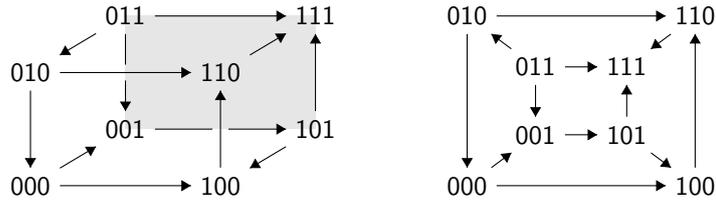
\begin{figure}[t]
\begin{center}
\begin{tikzpicture}[x=2.5cm,y=1.5cm,baseline=(current bounding box.center)]
 \foreach \x in {0,1}{
  \foreach \y in {0,1}{
    \foreach \z in {0,1}{
      \pgfmathsetmacro\xcord{1.0*\x+0.5*\z}
      \pgfmathsetmacro\ycord{1.0*\y+0.5*\z}
    \node (N\x\y\z) at (\xcord, \ycord) {$\x\y\z$}; 
 }}}
 
 \draw[->,endarrow] (N000) to node {} (N100);
 \draw[->,endarrow] (N011) to node {} (N111);
 \draw[<-,endarrow] (N010) to node {} (N000);
 \draw[<-,endarrow] (N011) to node {} (N001);
 \draw[->,endarrow] (N101) to node {} (N111);
 \draw[->,endarrow] (N000) to node {} (N001);
 \draw[<-,endarrow] (N011) to node {} (N010);
 \draw[<-,endarrow] (N101) to node {} (N100);
 \draw[->,endarrow] (N110) to node {} (N111);
 \draw[->,endarrow] (N001) to node {} (N101);

 \draw[-,color=white,line width=6pt] (N010) to node {} (N110);
 \draw[->,endarrow] (N010) to node {} (N110);
 \draw[-,color=white,line width=6pt] (N100) to node {} (N110);
 \draw[->,endarrow] (N100) to node {} (N110);
 \fill [black,opacity=0.1] (N001) rectangle (N111); 
 \end{tikzpicture}
\hspace*{1cm}
\begin{tikzpicture}[x=.6cm,y=.45cm,baseline=(current bounding box.center)]
 \foreach \x in {0,1}{
  \foreach \y in {0,1}{
   \foreach \z in {0,1}{
    \node (N\x\y\z) at({3*\x*(1-\z)-1.5*(1-\z)+2*\x},
                       {3*\y*(1-\z)-1.5*(1-\z)+2*\y}) {$\x\y\z$}; 
 }}}
 
 \draw[->,endarrow] (N000) to node {} (N100);
 \draw[->,endarrow] (N011) to node {} (N111);
 \draw[<-,endarrow] (N010) to node {} (N000);
 \draw[<-,endarrow] (N011) to node {} (N001);
 \draw[->,endarrow] (N101) to node {} (N111);
 \draw[->,endarrow] (N000) to node {} (N001);
 \draw[<-,endarrow] (N011) to node {} (N010);
 \draw[<-,endarrow] (N101) to node {} (N100);
 \draw[->,endarrow] (N110) to node {} (N111);
 \draw[->,endarrow] (N001) to node {} (N101);
 \draw[->,endarrow] (N010) to node {} (N110);
 \draw[->,endarrow] (N100) to node {} (N110);
 
 \end{tikzpicture}
 \caption{The $3$-dimensional twisted cube using parallel and
   perspective projections. On the left, the lid (i.e.\ the last face
   which can be recovered by filling) is shaded. On the right, this
   face is the small middle square. The lid can be seen as the
   composite of the other faces.}~\label{fig:3-cube-tw}
\end{center}
\end{figure}

\begin{figure}[t]
\begin{center}
  \begin{tikzpicture}[x=0.7cm,y=0.55cm]
    \foreach \x in {0,1}{
     \foreach \y in {0,1}{
      \foreach \z in {0,1}{
       \foreach \w in {0,1}{
        \pgfmathsetmacro\xcof{1.1}
        \pgfmathsetmacro\ycof{1}
        \pgfmathsetmacro\zxcof{0.3}
        \pgfmathsetmacro\zycof{0.4}
        \pgfmathsetmacro\wcof{2}
        
        \pgfmathsetmacro\xsign{2*\x - 1}
        \pgfmathsetmacro\ysign{2*\y - 1}
        \pgfmathsetmacro\zsign{2*\z - 1}
        
        \pgfmathsetmacro\wmult{\w * \wcof + 1}

        \pgfmathsetmacro\xsum{(\xcof * \xsign) + (\zxcof * \zsign)}
        \pgfmathsetmacro\ysum{(\ycof * \ysign) + (\zycof * \zsign)}
        
        \node (M\x\y\z\w) at (4.4 * \x + 2.2 * \z - 1.1 * \w
                             ,4.4 * \y + 1.1 * \z + 2.2 * \w)
                             {\tiny{$\x\y\z\w$}};
        \node (N\x\y\z\w) at (\wmult * \xsum + 13, \wmult * \ysum + 4)
                             {\tiny{$\x\y\z\w$}};
      }}}}

    \foreach \x in {0,1}{
     \foreach \y in {0,1}{
      \foreach \z in {0,1}{
        \draw[->,endarrow] (M0\x\y\z) -- (M1\x\y\z);
        \draw[->,midarrow] (N0\x\y\z) -- (N1\x\y\z);
        \pgfmathsetmacro\trg{\x}
        \pgfmathtruncatemacro\src{1 - \trg}
        \draw[->,endarrow] (M\x\src\y\z) -- (M\x\trg\y\z); 
        \draw[->,midarrow] (N\x\src\y\z) -- (N\x\trg\y\z); 
        \pgfmathtruncatemacro\trg{0.5*((2*\x - 1) * (2*\y - 1) + 1)}
        \pgfmathtruncatemacro\src{1 - \trg}
        \draw[->,endarrow] (M\x\y\src\z) -- (M\x\y\trg\z);
        \draw[->,midarrow] (N\x\y\src\z) -- (N\x\y\trg\z);
        \pgfmathtruncatemacro\trg{0.5*((2*\x - 1) * (2*\y - 1) *
                                  (2*\z - 1) + 1)}
        \pgfmathtruncatemacro\src{1 - \trg}
        \draw[->,endarrow] (M\x\y\z\src) -- (M\x\y\z\trg); 
        \draw[->,midarrow] (N\x\y\z\src) -- (N\x\y\z\trg); 
      }}}

  \fill [black,opacity=0.1] (-1.1, 2.2) -- (3.3, 2.2) -- (5.5, 3.3) --
                              (5.5, 7.7) -- (1.1, 7.7) -- (-1.1, 6.6) -- cycle; 
  \end{tikzpicture}
\end{center}
\caption{The $4$-dimensional twisted cube using parallel and
  perspective projections. The lid is shadowed on the left. It is the
  biggest cube on the right.}~\label{fig:4-cube-tw}
\end{figure}
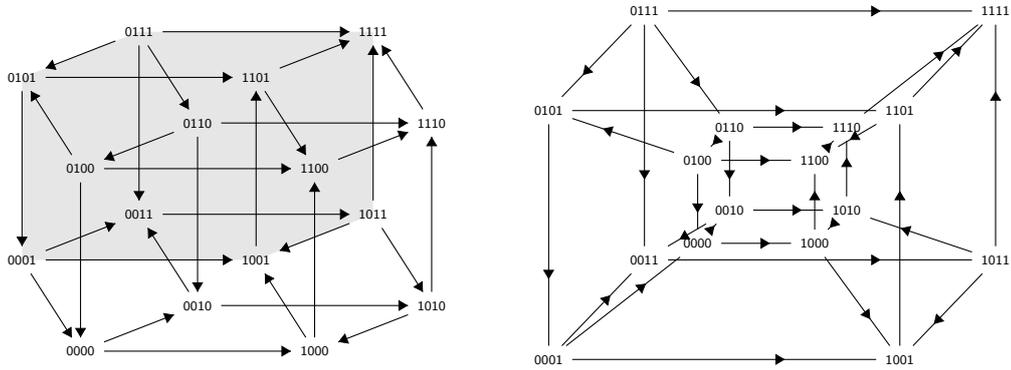

Twisted cubes do not only remove the discussed source of
invertibility, but they also change the way we view composition of
morphisms.  The filling of a ``standard'' square can be interpreted as
saying that the composition of two edges equals the composition of the
other two edges, and if we want to see the lid as the composite of the
three other edges, then one has to be inverted.  In contrast, in the
twisted square, the lid can be seen directly as the single composite
of the three other edges.  The right half of \cref{fig:3-cube-tw}
shows the projection of the twisted $3$-cube, and the smallest square
(011, 001, 101, 111) is the lid.  As for the square, this lid should
be seen as the composite of the other (here five) faces.  Intuitively,
one starts with the biggest square, composes it with the top and the
bottom squares, then with the left and the right square, and thus
arrives at the smallest square.  \Cref{fig:4-cube-tw} shows the
analogous situation for the $4$-dimensional twisted cube, where one
starts with the inner $3$-cube, then extends to the front and the
back, to the top and the bottom, and finally to the left and the
right.

The ``twisting'' pattern also appears in the \emph{twisted arrow
  category}~\cite{TwistedArrowCategory}, also known as the
\emph{category of factorisations}~\cite{CategoryOfFactorizations}.
However, it is unclear how to generalise this idea to more than
squares; it has been developed to solve a different problem.

In the main body of the paper, we first introduce the framework of
graph morphisms for standard (non-twisted) cubes.  We consider the
properties of meet/join and dimension preservation of graph morphisms,
and conclude that both of these are suitable refinements to ensure
that the category of graph morphisms matches the BCH cube category. The
proof of this is the main result of \cref{sec:standardcubes}.  We use
this development to introduce and examine \emph{twisted cubes} in
\cref{sec:twistedCubes}.  We will see that they have many
characteristic properties that standard cubes are lacking.  Some of
them, such as a Hamiltonian path through the cube and the fact that
vertices are totally ordered,
are familiar from simplicial structures but
not from cubical ones.  Another interesting feature, neither familiar
from cubical nor from simplicial but from globular structures, is
that surjective maps are unique (i.e.\ there is only one way to
degenerate a twisted cube).  These and other observations allow us to
define a further representation of the category of twisted cubes which
does not make use of graphs.

\subparagraph*{Setting} We use a standard version of Martin-L\"of's
dependent type theory as our meta-language.  We assume function
extensionality, but we do not require other axioms or features since
we mostly work with finite sets, which are extremely well-behaved by
default.  In particular, it does not matter for us whether UIP/Axiom K
is assumed or not, and the development would be identical in
extensional dependent type theory.

\subparagraph*{Summary of Contributions}
Our main contributions are as follows:
\begin{itemize}
 \item We give several alternative but equivalent presentations of the
   BCH cube category.
 \item We introduce \emph{twisted cubes}, a variation of the BCH cube
   category which allows for filling conditions without built-in
   invertibility.
 \item We show several results about twisted cubes.  These include
   connections to simplices (a unique Humiliation path and the
   property of being a Reedy category) and to globes (unique
   surjective maps and degeneracies).
\end{itemize}

\section{A Standard Cube Category}~\label{sec:standardcubes}

In this section, we discuss various representations of the cube
category $\shcube$.  This category was used by Bezem, Coquand, and
Huber to present a constructive model of
univalence~\cite{bezem_et_al:LIPIcs:2014:4628}.  In
\cref{sec:twistedCubes}, we will see how minimal modifications lead to
a category of twisted cubes.

Keeping in mind that we use type theory as the language in which the
results are presented (i.e.\ as our meta-theory), we use the following
notations: $\N$ are the natural numbers, including $0$.  For $n : \N$,
the set $\finset n$ is the finite set with elements
$\{0, 1, \ldots, n-1\}$.  In particular, $\finset 2$ is the set of
booleans.  As usual, ${\finset n}^{\finset m}$ is simply the function
set $\finset m \to \finset n$.  We denote elements of
${\finset 2}^{\finset n}$ by binary sequences as in
$0 \cdot 1 \cdot 1 \cdot 0$.  This means such a function $f$ is
denoted by $f(0) \cdot f(1) \cdot f(2)\ldots f(n-1)$.  If there is no
risk of confusion, we omit the $\cdot$ and simply use juxtaposition as
in $0110$.

In several situations, we want to consider a type of functions into a
coproduct which is injective ``on the \emph{left} part of the
codomain''.  To make this precise, we introduce a notation:
\begin{definition}[$\tolinj$]~\label{def:inj-left} Assume $A$, $B$,
  and $C$ are given types.  For a function $f : A \to (B+C)$, we say
  that $f$ is \emph{injective on the left part} if
 \begin{equation}
   \linj(f) \defeq \Pi(x,y : A, z : B). (f(x) = \inl(z)) \to (f(y) = \inl(z)) \to x = y.
 \end{equation}
 We write the type of functions which are injective on the left part as
 \begin{equation}
  (A \tolinj B+C) \defeq \Sigma (f: A \to (B+C)). \linj(f).
 \end{equation}
\end{definition}

In the next lemma, a function $f : A \to B + \finset 1$ is called a
\emph{partial function}, with $\finset 1$ being the ``undefined''
part.\footnote{%
  Technically, these are of course only the partial functions from $A$
  to $B$ with decidable support. Since we only work with finite types,
  it is not surprising that we only need to consider the decidable
  case.}  The following simple but useful (and well-known) result will
be necessary.  It could be formulated in higher generality, but a
version which is sufficient for us is this:

\begin{lemma}~\label{lem:inj-partial-sym} Given $m,n : \N$, injective
  partial functions from $\finset m$ to $\finset n$ are in bijection
  with injective partial functions from $\finset n$ to $\finset m$.
  In other words, we have an equivalence
\begin{equation}
  \left(\finset m \tolinj \finset n + \finset 1\right) \simeq \left(\finset n \tolinj \finset m + \finset 1\right).
\end{equation}
\end{lemma}
\begin{proof}
  The equivalence can be constructed directly. Given an
  $f : \finset m \tolinj \finset n + \finset 1$, we have to construct
  a function $g : \finset n \tolinj \finset m + \finset 1$.  For
  $i : \finset n$, we can decide whether there is a $k$ such that
  $f(k) = \inl(i)$.  If so, then this $k$ is unique due to
  injectivity, and we set $g(i) \defeq \inl(k)$; otherwise, we set
  $g(i) \defeq \inr(0)$.  Checking that this is an equivalence is
  routine.
\end{proof}

The presentation of the cube category in question that we start with
is the one given by Bezem, Coquand, and
Huber~\cite{bezem_et_al:LIPIcs:2014:4628} (which is the same as in
Huber's PhD thesis~\cite{simon:thesis}).  Since it is sufficient for
our purposes, we use a skeletal variation: our objects are not finite
sets but rather natural numbers.

\begin{definition}[category
  $\shcube$~\cite{bezem_et_al:LIPIcs:2014:4628,simon:thesis}]~\label{def:shcube}
  The category $\shcube$ has natural numbers as objects and, for
  $m, n : \N$, a morphism in $\shcube(m,n)$ is a function
  $f : \finset m \to \finset n + \finset 2$ which is injective on the
  $\finset n$-part.  In type-theoretic notation:
 \begin{align}
  & \obj{\shcube} \defeq \N & \mor{\shcube}(m,n) \defeq \finset m \tolinj \finset n + \finset 2
 \end{align}
 Composition $g \circ f$ is defined to be the set-theoretic
 composition $(g + \mathsf{id}_2) \circ f$.
\end{definition}

What we will need is the opposite of this category, $\shcubeop$.
While the above definition is short and abstract, a description close
to the intuitive idea of cubes is helpful for our later developments.
Let us consider \emph{graphs} $G = (V, E)$ of nodes (vertices) and
edges, where $V$ is a set with decidable equality and $E$ is a subset
of $V \times V$.  A standard way to implement this is to let $E$ be a
family of ``mere propositions''\footnote{Recall that a \emph{mere
    proposition}, or a \emph{subsingleton}, is a type with at most one
  element.}, indexed twice over $V$.  However, we write $(s,t) : E$
for $E(s,t)$ and assume that $E$ is given in the ``total space''
formulation.  Furthermore, in our cases $E$ will always be a
\emph{decidable} subset.

$E$ being a subset means that our graphs do \emph{not} have multiple
parallel edges, i.e.\ for any pair of vertices, there is at most one
edge between them, and it is decidable whether there is an edge
between two given vertices.

Given a graph, we construct a new graph as follows.  Note that the
``total space'' of the edges of the new graph is $E + E + V$, but in
order to make clear which vertices these new edges connect, we use
``set theory style'' notation:
\begin{definition}~\label{def:ord-graph-iter} Given $G = (V, E)$, the
  \emph{graph-prism} of $G$, denoted as \\
  $\graphprism \; (G) \defeq (\graphprism \; (V), \graphprism \; (E))$
  is another graph where
  \begin{align}
    \graphprism \; (V) \; \defeq \; & \bool \times V
    \\
    \graphprism \; (E) \; \defeq \; & \{ \;(\;(0, \;s), \;(0, t)\;)\;\; | \;\;(s, t) : E \} \label{eq:nontwisted-prechange}
    \\
    \cup \; & \{ \;(\;(1, \;s), \;(1, t)\;)\;\; | \;\;(s, t) : E \}
    \\
    \cup \; & \{ \;(\;(0, \;v), \;(1, v)\;)\;\; | \;\;v : V \}. 
  \end{align}
\end{definition}

This allows us to define the standard cube as a graph:\footnote{Most
  of graphs in this paper are reflexive graphs to support degeneracies
  as graph morphisms.}
\begin{definition}~\label{def:ord-cube-graph-rec}
  Given $n : \N$, the standard cube $\cube n$ is defined as follows:
  \begin{alignat}{2}
    \cube 0 \; \defeq\; (\finset 1, \; \{(0, 0)\})\qquad\qquad\qquad\qquad\qquad
    \cube {n + 1} \;\defeq\; \graphprism \; (\cube n)
  \end{alignat}
\end{definition}
Another way of defining $\cube n$, without recursion, is the
following.  Here, we give the ``total space'' of edges
$\mathsf{edges}(\cube n)$ together with functions
$\mathsf{src}, \mathsf{trg} : \mathsf{edges}(\cube n) \to
\mathsf{nodes}(\cube n)$:
\begin{definition}~\label{def:ord-cube-graph-nonrec} In the following,
  our convention is that $\finset {-1}$ is empty (i.e.\ the same as
  $\finset 0$):
 \begin{align}
  & \mathsf{nodes}(\cube n) && \defeq && \bool^{\finset n} \label{eq:def-cube-nonrec-nodes} \\
  & \mathsf{edges}(\cube n) && \defeq && \bool^{\finset n} 
     + \left(\finset n \times \bool^{\finset{n-1}} \right) \label{eq:def-cube-nonrec-edges} \\
  & \mathsf{src}(\inl(v)) \quad \defeq \quad \mathsf{trg}(\inl(v)) && \defeq && v \\
  & \mathsf{src}(\inr(i, x_0x_1\ldots x_{n-2})) && \defeq && x_0x_1\ldots x_{i-1} 0 x_i \ldots x_{n-2} \\
  & \mathsf{trg}(\inr(i, x_0x_1\ldots x_{n-2})) && \defeq && x_0x_1\ldots x_{i-1} 1 x_i \ldots x_{n-2}
 \end{align}
\end{definition}

\begin{figure}[b] \centering
\begin{tikzpicture}[x=1.2cm,y=1.2cm,baseline=(current bounding box.center)]
\node (N) at (0,0.5) {$\epsilon$}; 
\node (N0) at (2,0.5) {$0$}; 
\node (N1) at (3,0.5) {$1$}; 
\draw[->,endarrow] (N0) to node[above,sloped]
 {\tiny{$\langle 0, \epsilon \rangle$}} (N1);

 \foreach \x in {0,1}{
  \foreach \y in {0,1}{
     \node (N\x\y) at (\x + 5,\y) {$\x\y$}; 
 }}
\draw[->,endarrow] (N00) to node[above,sloped]
 {\tiny{$\langle 0, 0 \rangle$}} (N10);
\draw[->,endarrow] (N01) to node[above,sloped]
 {\tiny{$\langle 0, 1 \rangle$}} (N11);
\draw[->,endarrow] (N00) to node[above,sloped]
 {\tiny{$\langle 1, 0 \rangle$}} (N01);
\draw[->,endarrow] (N10) to node[above,sloped]
 {\tiny{$\langle 1, 1 \rangle$}} (N11);
  
 \foreach \x in {0,1}{
  \foreach \y in {0,1}{
   \foreach \z in {0,1}{
     \node (N\x\y\z) at (1.5*\x*\z-0.75*\z+\x + 9,
                         1.5*\y*\z-0.75*\z+\y) {$\x\y\z$}; 
 }}}

    \foreach \x in {0,1}{
     \foreach \y in {0,1}{
       \draw[->,endarrow] (N0\x\y) to node[above,sloped]
        {\tiny{$\langle 0, \x\y \rangle$}} (N1\x\y);
       \draw[->,endarrow] (N\x0\y) to node[above,sloped]
        {\tiny{$\langle 1, \x\y \rangle$}} (N\x1\y); 
       \draw[->,endarrow] (N\x\y0) to node[above,sloped]
        {\tiny{$\langle 2, \x\y \rangle$}} (N\x\y1);
         }}
 \end{tikzpicture}
 \caption{An illustration of $\cube n$ for $n \leqslant 3$. The labels
   on the vertices and edges are in accordance
   with~\eqref{eq:def-cube-nonrec-nodes}
   and~\eqref{eq:def-cube-nonrec-edges}. The identity loops are
   omitted. This allows us to unambiguously hide the constructor
   $\inr{}$ as well. }~\label{fig:ord-cube-graph-123}
\end{figure}
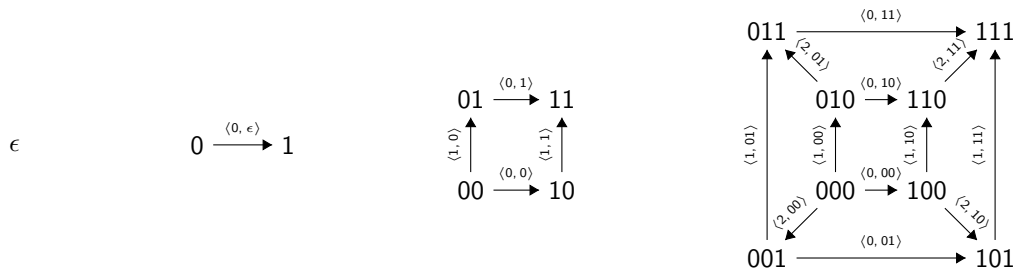

The number of total edges in~\eqref{eq:def-cube-nonrec-edges} comes
from the following calculation.  We have $n$ dimension, thus $2^n$
nodes, which come with self-loops giving rise to the summand
$\bool^{\finset n}$.  For ever node, we further have an edge in each
dimension.  Avoiding double counting, this gives the summand
$\finset n \times \bool^{\finset{n-1}}$.
\Cref{fig:ord-cube-graph-123} shows drawings for $\cube 0$ to
$\cube 3$.

\begin{lemma}
\Cref{def:ord-cube-graph-rec} and \cref{def:ord-cube-graph-nonrec} 
define isomorphic graph structures. \qed{}
\end{lemma}
This observation allows us to use whichever is more convenient in any
given situation.

A \emph{graph morphism} from $G = (V,E)$ to $G' = (V',E')$ is, as
usual, a function between the node types which preserves the edges:
\begin{equation}
 \graphhom{(V,E)}{(V',E')} \defeq \Sigma (f : V \to V'). \Pi(v_0, v_1 : V). E(v_0,v_1) \to E'(f(v_0),f(v_1))
\end{equation}

We can now consider the following category:
\begin{definition}[category $\graphcube$]~\label{def:ord-graphs-def}
  The category $\graphcube$ has natural numbers as objects. \\ A
  morphism between $m$ and $n$ is a graph morphism from $\cube m$ to
  $\cube n$, as in:
 \begin{align}
  & \obj{\graphcube} \defeq \N
  & \mor{\graphcube}(m,n) \defeq \graphhom{\cube m}{\cube n}
 \end{align}
 Composition is composition of graph morphisms.
\end{definition}

The category $\graphcube$ has more morphisms than $\shcubeop$.  One
example would be the morphism in $\graphhom{\cube 2}{\cube 1}$ which
maps the three nodes $00$, $01$, $10$ all to $0$ and $11$ to $1$.
Another example is the morphism which maps $00$ to $0$, and $01$,
$10$, $11$ all to $1$.  Both of these morphisms do not have analogues
in $\shcubeop$.  In other words, $\graphcube$ has \emph{connections}.
Since, in the current paper, we are looking for alternative
definitions of the category $\shcubeop$, we refine the definition of
the morphisms in $\graphcube$ to resolve the mismatch.  Let us
formulate the following auxiliary definitions.

\begin{definition}[free preorder on a graph]~\label{def:freepreorder}
  For a given graph $G = (V,E)$, we write $G^* = (V, E^*)$ for the
  free preorder generated by it.  $G^*$ has $V$ as objects and, for
  $v,u : V$, we have $v \leqslant u$ if there is a chain of edges
  starting in $v$ and ending in $u$.
 
  When talking about nodes in $G$, we borrow the notions of
  \emph{meet} (product) and \emph{join} (coproduct) from preorders.
  If they exist in $G^*$, we write them as $v \meet u$ and
  $v \join u$.
\end{definition}

It is easy to see that, in the case of $\cube n$, all meets and joins
exist and can be calculated directly: From the programming
perspective, they correspond to the bitwise operators $'\texttt{\&}'$
and $'\texttt{|}'$.  Thus, when talking about $\cube n$, we can view
$\meet$ and $\join$ as actual functions calculating the binary meet
and join:
\begin{equation}
 \meet, \join : V \times V \to V
\end{equation}
Given a graph morphism $g : \graphhom{\cube m}{\cube n}$, it is easy
to define what it means that it preserves binary meets resp.\ joins:
\begin{align}
 & \presmeet(g) \defeq \Pi (u,v : \bool^{\finset m}). g(u \meet v) = g(u) \meet g(v) \\
 & \presjoin(g) \defeq \Pi (u,v : \bool^{\finset m}). g(u \join v) = g(u) \join g(v)
\end{align}
Note that preserving meets and joins is a property (a ``mere
proposition'') of morphisms.  For general morphisms between graphs
which might not have all meets or joins, the definition is more subtle
but still straightforward; one can always define the property of
\emph{being a meet (join)} and then say that any vertex which has this
property is mapped to one which also has it.  We omit the precise
type-theoretic formulation.

The two mentioned examples of morphisms which are ``too much'' in
$\graphcube$ do not preserve binary meets resp.\ joins.

\begin{definition}[category $\graphmeet$]~\label{def:graphmeet} The
  category $\graphmeet$ has $\N$ as objects and, as morphisms, graph
  morphisms between standard cubes which preserve meets and joins
  ($\mathsf{cont}$ for \emph{continuous}):
 \begin{align}
  & \obj{\graphmeet} \defeq \N \\
   & \mor{\graphmeet}(m,n) \defeq \Sigma (g : \graphhom{\cube m}{\cube n}).
     \presmeet(g) \times \presjoin(g)
 \end{align}
\end{definition}

This gives us a category which is indeed equivalent (in fact
isomorphic) to $\shcubeop$:

\begin{theorem}~\label{thm:shcube-meetjoinpres} The categories
  $\shcubeop$ and $\graphmeet$ are isomorphic.  The isomorphism on the
  object part is the identity, i.e.\ the equivalence is given by a
  family $e$ as in:
 \begin{equation}
  e : \Pi(m,n : \N). \mor{\shcubeop}(m,n) \simeq \mor{\graphmeet}(m,n).
 \end{equation}
\end{theorem}

Before giving a proof, we formulate the following:
\begin{lemma}~\label{lem:graphmeet-reduce} Consider the full subgraph
  of $\cube n$ which has exactly $(n+1)$ vertices, namely the
  ``origin'' $00\ldots 0$ and the ``base vectors'' which have exactly
  one $1$.  We call this subgraph $B_n$, where the $B$ stands for
  ``base'', and it comes with the inclusion
  $i : B_n \hookrightarrow \cube n$.  For any $m$, ``forgetting'' the
  property of preserving the joins and composing with $i$ as in
 \begin{equation}
  \lambda g. i \circ (\mathsf{proj}_1(g)) : \; \left(\Sigma (g : \graphhom{\cube m}{\cube n}. \presjoin(g)\right) \; \to \; \graphhom{B_m}{\cube n}
 \end{equation}
 is an equivalence.  Moreover, $g$ preserves meets if and only if
 $i \circ (\mathsf{proj}_1(g))$ does.
\end{lemma}
\begin{minipage}{.75\textwidth}
\begin{proof}
  The only binary joins that $B_m$ has are trivial, so every morphism
  $\graphhom{B_m}{\cube n}$ is join-preserving. Thus, the first claim
  of the lemma is that every such morphism can be extended in a unique
  way as shown in the diagram to the right.  Every node of $\cube m$
  which is not in $B_m$, i.e.\ every node which is not the origin or a
  base vector, can be written as a join of base vectors.  Since we
  need to preserve joins, it is therefore determined where the node
  has to be sent to.  The map defined in this way preserves all binary
  joins, and it preserves binary meets if and only if the input does.
\end{proof}
\end{minipage}
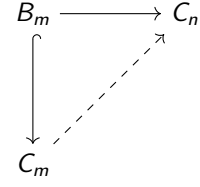
\hfill
\begin{minipage}{.20\textwidth}
\begin{tikzpicture}[x=2cm,y=-2cm,baseline=(current bounding box.center)]
 \node (Gn) at (0,0) {$B_m$};
 \node (Cm) at (1,0) {$\cube n$};
 \node (Cn) at (0,1) {$\cube m$};

 \draw[->] (Gn) to node {} (Cm);
 \draw[right hook->] (Gn) to node {} (Cn);
 \draw[->,dashed] (Cn) to node {} (Cm);
\end{tikzpicture}
\end{minipage}

\begin{proof}[Proof of \cref{thm:shcube-meetjoinpres}]
  We first give the overview of the argument as a chain of
  equivalences, then we justify each step [S1 -- S5].
\begin{alignat*}{4}
  && \; && \;\; & \mor{\graphmeet}(m,n) \\
   &&& \equiv && \Sigma (g : \graphhom{\cube m}{\cube n}). \presmeet(g) \times \presjoin(g) \\
  &\step 1 && \simeq && \Sigma (g : \graphhom{B_m}{\cube n}). \presmeet(g) \\
  &\step 2 && \simeq && \Sigma(z : {\finset 2}^{\finset n}, d : \finset m \tolinj \finset n + \finset 1). \Pi(i:\finset m, j : \finset n). (d(i) = \inl(j)) \to (z(j) = 0) \\
  &\step 3 && \simeq && \Sigma(z : {\finset 2}^{\finset n}, e : \finset n \tolinj \finset m + \finset 1). \Pi(i:\finset m, j : \finset n). (e(j) = \inl(i)) \to (z(j) = 0) \\
  &\step 4 && \simeq && \Sigma(z : {\finset 2}^{\finset n}, e : \finset n \to (\finset m + \finset 1)). \linj(e) \times \Pi(i:\finset m, j : \finset n). (e(j) = \inl(i)) \to (z(j) = 0) \\
  &\step 5 && \simeq && \Sigma \big(\alpha : \Pi(j : \finset n). \Sigma(e : \finset m + \finset 1, z : \finset 2). \Pi(i : \finset m).(e = \inl(i)) \to z = 0\big). \linj(\mathsf{proj}_1 \circ \alpha) \\
  &\step 6 && \simeq && \Sigma \big(\alpha : \Pi(j : \finset n). \finset m + \finset 2\big). \linj(\alpha) \\
   &&& \equiv && \mor{\shcubeop}(m,n) 
\end{alignat*}

Step 1 holds by \cref{lem:graphmeet-reduce}.  Let us look at Step 2.
Giving a graph homomorphism between $B_m$ and $\cube n$ corresponds to
choosing where the origin is mapped to, and choosing where each
(non-trivial) edge of $B_m$ is mapped to.  For the origin, we use the
component $z : {\finset 2}^{\finset n}$.  There are $m$ non-trivial
edges in $B_m$, and $z$ is an endpoint (or starting point) of $n$
non-trivial edges and one trivial edge in $\cube n$.  This gives us up
to $\finset m \to \finset n + \finset 1$ possible functions, but since
we only consider meet-preserving morphisms, every function needs to be
injective on the left part, leading to
$d : \finset m \tolinj \finset n + \finset 1$.  Moreover, if
$d(i) = \inl(j)$ for some $i,j$, then the image of the origin must be
the \emph{starting point} of the edge in dimension $j$, i.e.\
$z(j) = 0$.  Step 3 is an application of \cref{lem:inj-partial-sym}
(it essentially swaps the roles of $m$ and $n$).  Step 4 only unfolds
the definition of $\tolinj$.

In Step 5, the usual distributivity between $\Sigma$ and $\Pi$ (under
the propositions-as-types view referred to as the ``axiom of choice'')
is used: $z$, $e$, and the unnamed last component can all be seen as
(dependent) functions with domain $\finset n$.  The dependent function
$\alpha$ combines them into a single dependent function with domain
$\finset n$ and a codomain that consists of multiple components which,
again, are called $e$, $z$, with the last one being unnamed.  Only the
component expressing the ``injectivity on the left part''-property
cannot be seen as a function in $\finset n$.  In Step 6, we massage
the codomain of $\alpha$: We have $e : \finset m + \finset 1$ and also
$z : \finset 2$, but the condition says that $z$ is determined unless
$e = \inr(0)$; thus, the type is equivalent to
$\finset m + \finset 2$.

We omit the calculation which shows that the constructed equivalence
preserves composition of morphisms in the categories.
\end{proof}

In \cref{sec:twistedCubes}, we will switch from standard cubes to
twisted cubes.  The directions of some edges will be reversed.  It is
therefore an advantage to formulate a condition similar to the one
about meets and joins without referring to the direction of edges.
This is indeed possible:

\begin{definition}[dimension preserving morphisms; category
  $\graphdim$]~\label{def:ord-dim-pres} Given the standard cube
  $\cube n$, where we use the non-recursive definition as in
  \cref{def:ord-cube-graph-nonrec}, the \emph{dimension} of an edge is
  defined as follows:
 \begin{align}
  & \dimens : \mathsf{edges}(\cube n) \to \finset n + \finset 1
  && \dimens (\inl(v)) \qquad\quad\;\;\, \defeq \inr(0) \\
  &&& \dimens (\inr(i, x_0\ldots x_{n-2}) \defeq \inl(i)
 \end{align}
 We say that a morphism $f : \graphhom{\cube m}{\cube n}$ is
 \emph{dimension-preserving} if $f$ maps edges of the same dimension
 to edges of the same dimension,
 \begin{equation} \label{eq:dim-preserving}
  \dimpres(f) \defeq \Pi(e_1,e_2 : \mathsf{edges}(\cube n)). (\dimens(e_1) = \dimens(e_2)) \to (\dimens(f(e_1)) = \dimens(f(e_2))).
 \end{equation}
 The category $\graphdim$ makes use of these concepts:
 \begin{align}
  & \mathsf{obj}(\graphdim) \defeq \N
  & \mor{\graphdim}(m,n) \defeq \Sigma (g : \graphhom{\cube m}{\cube n}). \dimpres(g)
 \end{align}
\end{definition}

As $\presmeet(g)$ and $\presjoin(g)$, preserving the dimension as
in~\eqref{eq:dim-preserving} is a proposition in the sense of homotopy
type theory (has at most one proof).

\begin{remark}~\label{rm:dim-inj-derivable} For a graph morphism $f$
  as in the definition above, the following condition says that $f$ is
  ``injective on dimensions'' (on the non-trivial part):
 \begin{align*}
  & \diminj(f) \defeq \Pi(e_1,e_2 : \mathsf{edges}(\cube m), j : \finset n) . 
    \big( \dimens(f(e_1)) = \inl(j) \times \dimens(f(e_2)) = \inl(j) \big) \\
  &  \hspace*{6cm} \to (\dimens(e_1) = \dimens(e_2)).
 \end{align*}
 However, note that this follows directly from $\dimpres(f)$: Assume
 $e_1, e_2$ are edges such that $\dimens(f(e_1))$ and
 $\dimens(f(e_2))$ are equal and non-trivial.  If $e_1$ and $e_2$ are
 not ``parallel'' (i.e.\ not in the same dimension), then we can find
 $e_1'$ in the same dimension as $e_1$ such that $e_1'$ and $e_2$ are
 adjacent (i.e.\ the endpoint of one is the starting point of the
 other). It is clear that $f(e_1')$ and $f(e_2)$ cannot go into the
 same non-trivial direction, since we can only go one step into a
 given direction before going back.
 \end{remark}

The connection to meet- and join-preserving is given by the following result:
\begin{lemma}~\label{lem:join-meet-dim} A morphism
  $f : \graphhom{\cube m}{\cube n}$ is join-and-meet-preserving
  exactly if it is dimension-preserving.
\end{lemma}
\begin{proof}
  This follows easily by going via morphisms $\graphhom{B_m}{C_n}$ as
  in \cref{lem:graphmeet-reduce}.  The graph $B_m$ has exactly one
  edge for every non-trivial dimension, and the proof is analogous to
  the one of \cref{lem:graphmeet-reduce}.
\end{proof}

\begin{corollary}[Section summary]
  The categories $\shcubeop$, $\graphmeet$, and $\graphdim$ are
  isomorphic. \qed{}
\end{corollary}

\section{A Category of Twisted Cubes}~\label{sec:twistedCubes}

As discussed in the introduction, we build on our framework of graph
morphisms to define a category of \emph{twisted cubes}.  A variation
of \cref{def:ord-graph-iter} gives us these twisted cubes.  The
critical change can be seen in~\eqref{eq:twisted-change}, which should
be compared with~\eqref{eq:nontwisted-prechange}:

\begin{definition}~\label{def:tw-graph-iter} Given a graph
  $G = (V,E)$, the \emph{twisted graph-prism} of $G$, \\ denoted as
  $\twgraphprism \; (G) \defeq (\twgraphprism \; (V), \; \twgraphprism
  \; (E))$ is the graph defined by
  \begin{align}
    \twgraphprism \; (V) \; \defeq \; & \bool \times V  \label{eq:nodes-2n}
    \\
    \twgraphprism \; (E) \; \defeq \; & \{ \;(\;(0, \;t), \;(0, s)\;)\;\; | \;\;(s, t) : E \} \label{eq:twisted-change}
    \\
    \cup \; & \{ \;(\;(1, \;s), \;(1, t)\;)\;\; | \;\;(s, t) : E \}
    \\
    \cup \; & \{ \;(\;(0, \;v), \;(1, v)\;)\;\; | \;\;v : V \}. \label{eq:twisted-edges-from-nodes}
  \end{align}
\end{definition}

We then define:
\begin{definition}~\label{def:tw-cube-graph-rec}
  Given $n : \N$, the twisted cube $\twcube n$ is defined as follows:
  \begin{alignat}{2}
    \twcube 0 \; \defeq\; (\finset 1, \; \{(0, 0)\})\qquad\qquad\qquad\qquad\qquad
    \twcube {n + 1} \;\defeq\; \twgraphprism \; (\twcube n)
  \end{alignat}
\end{definition}

Alternatively, we can tweak \cref{def:ord-cube-graph-rec} to get a
non-recursive definition. As before, the convention is that
$\finset {-1}$ is empty.
\begin{definition}~\label{def:tw-cube-graph-nonrec}
 The non-recursive definition of $\twcube n$ is as follows:
 \begin{align}
  & \mathsf{nodes}(\twcube n) && \defeq && \bool^{\finset n} \\
  & \mathsf{edges}(\twcube n) && \defeq && \bool^{\finset n} 
     + \left(\finset n \times \bool^{\finset{n-1}} \right) \label{eq:def-tw-cube-nonrec-edges} \\
  & \mathsf{src}(\inl(v)) \quad \defeq \quad \mathsf{trg}(\inl(v)) && \defeq && v \\
  & \mathsf{src}(\inr(i, x_0x_1\ldots x_{n-2})) && \defeq && x_0x_1\ldots x_{i-1} \cdot b \cdot x_i \ldots x_{n-2} \\
  & \mathsf{trg}(\inr(i, x_0x_1\ldots x_{n-2})) && \defeq && x_0x_1\ldots x_{i-1} \cdot (1 - b) \cdot x_i \ldots x_{n-2}
 \end{align}
 where $b = 1$ if the total number of zeros in $x_0x_1\ldots x_{i-1}$
 is odd, and $b = 0$ otherwise.
\end{definition}
This means that an edge is reversed (compared to the standard cubes
discussed before) exactly if the number of zeros in dimensions that
come \emph{before} the edge is odd (note that the condition talks
about $x_{i-1}$, not $x_{n-2}$).  The twisted cubes of dimension up to
$3$ are illustrated in \cref{fig:twisted-cube-graph-0123}; see also
\cref{fig:3-cube-tw,fig:4-cube-tw} in the introduction.

\begin{lemma}
  \Cref{def:tw-cube-graph-rec} and \cref{def:tw-cube-graph-nonrec}
  define isomorphic graph structures. \qed{}
\end{lemma}

\begin{figure}[b] \centering
\centering
\begin{tikzpicture}[x=1.2cm,y=1.2cm,baseline=(current bounding box.center)]
\node (N) at (0,0.5) {$\epsilon$}; 
\node (N0) at (2,0.5) {$0$}; 
\node (N1) at (3,0.5) {$1$}; 
\draw[->,endarrow] (N0) to node[above,sloped]
 {\tiny{$\langle 0, \epsilon \rangle$}} (N1);

 \foreach \x in {0,1}{
  \foreach \y in {0,1}{
     \node (N\x\y) at (\x + 5,\y) {$\x\y$}; 
 }}
\draw[->,endarrow] (N00) to node[above,sloped]
 {\tiny{$\langle 0, 0 \rangle$}} (N10);
\draw[->,endarrow] (N01) to node[above,sloped]
 {\tiny{$\langle 0, 1 \rangle$}} (N11);
\draw[->,endarrow] (N01) to node[above,sloped]
 {\tiny{$\langle 1, 0 \rangle$}} (N00);
\draw[->,endarrow] (N10) to node[above,sloped]
 {\tiny{$\langle 1, 1 \rangle$}} (N11);
  
 \foreach \x in {0,1}{
  \foreach \y in {0,1}{
   \foreach \z in {0,1}{
     \node (N\x\y\z) at (1.5*\x*\z-0.75*\z+\x + 9,
                         1.5*\y*\z-0.75*\z+\y) {$\x\y\z$}; 
 }}}
 
    \foreach \x in {0,1}{
     \foreach \y in {0,1}{
        \draw[->,endarrow] (N0\x\y) to node[above,sloped]
         {\tiny{$\langle 0, \x\y \rangle$}} (N1\x\y);
        \pgfmathsetmacro\trg{\x}
        \pgfmathtruncatemacro\src{1 - \trg}
        \draw[->,endarrow] (N\x\src\y) to node[above,sloped]
         {\tiny{$\langle 1, \x\y \rangle$}} (N\x\trg\y); 
        \pgfmathtruncatemacro\trg{0.5*((2*\x - 1) * (2*\y - 1) + 1)}
        \pgfmathtruncatemacro\src{1 - \trg}
        \draw[->,endarrow] (N\x\y\src) to node[above,sloped]
         {\tiny{$\langle 2, \x\y \rangle$}} (N\x\y\trg);
         }}
 
 \end{tikzpicture}
 \caption{An illustration of $\twcube n$ where
   $n \leqslant 3$.}~\label{fig:twisted-cube-graph-0123}
\end{figure}

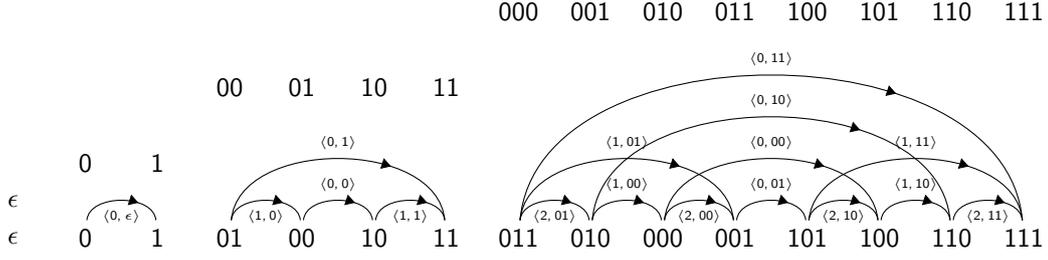
\begin{figure}[t]
    \begin{center}
        \begin{tikzpicture}[x=0.95cm,y=1.0cm]
        \def\bendparam{85}
        
        \node (N0) at (0 + 0, 0) {$\epsilon$};
        \node () at (0 + 0, 0.5) {$\epsilon$};
        
        \node (N0) at (1 + 0, 0) {$0$};
        \node (N1) at (1 + 1, 0) {$1$};
        \node () at (1 + 0, 1) {$0$};
        \node () at (1 + 1, 1) {$1$};
        
        \node (N01) at (3 + 0, 0) {$01$};
        \node (N00) at (3 + 1, 0) {$00$};
        \node (N10) at (3 + 2, 0) {$10$};
        \node (N11) at (3 + 3, 0) {$11$};
        \foreach \x in {0,1}{
            \foreach \y in {0,1}{
                \node () at (3 + 2*\x + \y, 2) {$\x\y$};
        }}
        
        \node (N011) at (7 + 0, 0) {$011$};
        \node (N010) at (7 + 1, 0) {$010$};
        \node (N000) at (7 + 2, 0) {$000$};
        \node (N001) at (7 + 3, 0) {$001$};
        \node (N101) at (7 + 4, 0) {$101$};
        \node (N100) at (7 + 5, 0) {$100$};
        \node (N110) at (7 + 6, 0) {$110$};
        \node (N111) at (7 + 7, 0) {$111$};
        \foreach \x in {0,1}{
            \foreach \y in {0,1}{
                \foreach \z in {0,1}{
                    \node () at (7 + 4*\x + 2*\y + \z, 3) {$\x\y\z$};
        }}}
        
        \draw[->,midarrow] (N0) to [bend left=\bendparam]
        node[below]{\tiny{$\langle 0, \epsilon \rangle$}} (N1);
        
        \draw[->,midarrow] (N00) to [bend left=\bendparam] 
        node[above]{\tiny{$\langle 0, 0 \rangle$}} (N10);
        \draw[->,midarrow] (N01) to [bend left=\bendparam]
        node[above]{\tiny{$\langle 0, 1 \rangle$}} (N11);
        \draw[->,midarrow] (N01) to [bend left=\bendparam] 
        node[below]{\tiny{$\langle 1, 0 \rangle$}} (N00);
        \draw[->,midarrow] (N10) to [bend left=\bendparam] 
        node[below]{\tiny{$\langle 1, 1 \rangle$}} (N11);
        
        \draw[->,midarrow] (N000) to [bend left=\bendparam]
        node[above]{\tiny{$\langle 0, 00 \rangle$}} (N100);
        \draw[->,midarrow] (N001) to [bend left=\bendparam] 
        node[above]{\tiny{$\langle 0, 01 \rangle$}} (N101);
        \draw[->,midarrow] (N010) to [bend left=\bendparam] 
        node[above]{\tiny{$\langle 0, 10 \rangle$}} (N110);
        \draw[->,midarrow] (N011) to [bend left=\bendparam] 
        node[above]{\tiny{$\langle 0, 11 \rangle$}} (N111);
        
        \draw[->,midarrow] (N010) to [bend left=\bendparam]
        node[above]{\tiny{$\langle 1, 00 \rangle$}} (N000);
        \draw[->,midarrow] (N011) to [bend left=\bendparam] 
        node[above]{\tiny{$\langle 1, 01 \rangle$}} (N001);
        \draw[->,midarrow] (N100) to [bend left=\bendparam] 
        node[above]{\tiny{$\langle 1, 10 \rangle$}} (N110);
        \draw[->,midarrow] (N101) to [bend left=\bendparam] 
        node[above]{\tiny{$\langle 1, 11 \rangle$}} (N111);
        
        \draw[->,midarrow] (N000) to [bend left=\bendparam]
        node[below]{\tiny{$\langle 2, 00 \rangle$}} (N001);
        \draw[->,midarrow] (N011) to [bend left=\bendparam] 
        node[below]{\tiny{$\langle 2, 01 \rangle$}} (N010);
        \draw[->,midarrow] (N101) to [bend left=\bendparam] 
        node[below]{\tiny{$\langle 2, 10 \rangle$}} (N100);
        \draw[->,midarrow] (N110) to [bend left=\bendparam] 
        node[below]{\tiny{$\langle 2, 11 \rangle$}} (N111); 
        \end{tikzpicture}
    \end{center}
    \caption{Linear drawings of the twisted cubes $\twcube 0$,
        $\twcube 1$, $\twcube 2$, and $\twcube 3$, demonstrating that the
        underlying preorders are total orders. The binary sequences on top
        are the values of $g_n$ from the proof of
        \cref{thm:total-order}. See also
        \cref{rm:positional-ordinal-binary-number}.}~\label{fig:rainbows}
\end{figure}

$\twcube n$ has an interesting property that the standard cube
$\cube n$ does not have: The induced preorder $\twcube n^*$ on the
vertices is a total order.  This observation was originally suggested
by Paolo Capriotti and Jakob von Raumer in a discussion with the first
author of this paper. Note that this observation should not be
misunderstood to mean that $\twcube n$ itself is uninteresting.  Its
edges give it a unique structure, as visualised in
\cref{fig:rainbows}.

The idea behind this result is that $\twgraphprism$ preserves the
property of having a preorder that is total.  To elaborate on this, if
$G^*$ is a total order, then ${(\twgraphprism \; G)}^*$ consists of
two copies of $G^*$, where the first copy is ``turned around''.  One
of the edges added in~\eqref{eq:twisted-edges-from-nodes} links the
largest node in the first copy to the smallest node in the second
copy, thus every element of the second copy is larger than all the
elements of the first copy.  In other words,
${(\twgraphprism \; G)}^*$ is the \emph{join} of the two copies.%
\footnote{\emph{Join} in the sense of the \emph{join of categories} \cite{lurie:higher-topoi}, which should not be confused with the join (coproduct) of objects in a preorder (cf.~\cref{def:freepreorder}).}

\begin{theorem}~\label{thm:total-order} For all $n : \N$, the preorder
  $\twcube n^*$ is isomorphic to the total order
  $({\finset 2}^{\finset n}, <)$.
\end{theorem}

Note that \cref{thm:total-order} is a property which one usually
expects for simplicial structures, but not for cubical ones.

\begin{remark}~\label{rm:positional-ordinal-binary-number} There are
  two binary numbers for each node in~\cref{fig:rainbows}.  The bottom
  one represents each node name according
  to~\cref{def:tw-cube-graph-nonrec} whereas the top one represents
  the total order of $\twcube 3$.  It is impossible to unify these two
  binary numbers for $n \geqslant 2$ since, for each edge $e$, the
  numbers $\mathsf{src}(e)$ and $\mathsf{src}(e)$ only differ by (at
  most) one single bit by~\cref{def:tw-cube-graph-nonrec}, while
  incrementing a binary number can flip more than one bit.
\end{remark}

Another related observation is that we can find a path from the
smallest vertex to the largest vertex of $\twcube n$ which respects
the direction of the edges, and which visits each vertex exactly
once. Recall that such a path is called a \emph{Hamiltonian path}.  We
record this:

\begin{theorem}~\label{thm:hamiltonian} For all $n : \N$, there is
  exactly one Hamiltonian path through $\twcube {n+1}$.  This path
  contains exactly one edge in the first dimension (i.e.\ the one
  which is added when going from $\twcube n$ to $\twcube {n+1}$).
  Moreover, this single edge in the new dimension connects the
  Hamiltonian paths through the two copies of $\twcube n$ of which
  $\twcube {n+1}$ consists by definition, cf.~\eqref{eq:nodes-2n}.
\end{theorem}
\begin{proof} [Proof of \cref{thm:total-order} and \cref{thm:hamiltonian}]
  As before, we denote elements of ${\finset 2}^{\finset n}$ as
  sequences such as $00101$ (binary representation with most
  significant bit first) or, for clarity, by
  $0 \cdot 0 \cdot 1 \cdot 0 \cdot 1$.  We use the endofunction
  $\negate$ on $\finset 2^{\finset n}$, which simply replaces each $0$
  in a sequence by a $1$ and vice versa; i.e.\ it sends the number $i$
  to $2^n - 1 - i$ (note that $\negate$ does not reverse the sequence,
  but the ordering on ${\finset 2}^{\finset n}$).
 
  Let us define endofunctions $f_n$ and $g_n$ on
  ${\finset 2}^{\finset n}$, by induction on $n$. Note that, at this
  point, we do not talk about graph morphisms but only about functions
  between sets.  The base cases of the induction are uniquely
  determined. We define $f$ and $g$ by
 \begin{align}
  & f_{n+1} (0 \cdot \vec x) \defeq 0 \cdot f_n(\negate(\vec x)) & & g_{n+1} (0 \cdot \vec x) \defeq 0 \cdot \negate(g_n(\vec x)) \label{eq:def-f-0}   \\
  & f_{n+1} (1 \cdot \vec x) \defeq 1 \cdot f_n(\vec x)  & & g_{n+1} (1 \cdot \vec x) \defeq 1 \cdot g_n(\vec x). \label{eq:def-f-1}
 \end{align}

 It is easy to calculate that, by induction, $f$ and $g$ are inverse
 to each other.  We want to show that they extend to morphisms between
 preorders,
 \begin{align}
  & \hat f_n : ({\finset 2}^{\finset n}, <) \to \twcube n^*
  & \hat g_n : \twcube n^* \to ({\finset 2}^{\finset n}, <).
 \end{align}
 To construct $\hat f_n$ \emph{and} the Hamiltonian path through the
 cube, it suffices to show: for $x,y : {\finset 2}^{\finset n}$ with
 $x+1=y$, we have an edge $f_n(x) \to f_n(y)$.
 
 We do induction on $n$. For $n = 0$, this is vacuously true (such
 $x,y$ do not exist).  For $n = n' + 1$, there are multiple cases:
 \begin{itemize}
 \item case $x = 0 \cdot x'$ and $y = 0 \cdot y'$: Then, the
   assumption gives us $x'+1=y'$ and we have to find an edge
   $0 \cdot f_n(\negate(x')) \to 0 \cdot f_n(\negate(y'))$.  Looking
   at \cref{def:tw-graph-iter}, we can get this if we have
   $f_n(\negate(y')) \to f_n(\negate(x'))$.  This holds by induction,
   since $\negate$ reverses the order which gives us
   $\negate(y')+1 = \negate(x')$.
 \item case $x = 1 \cdot x'$ and $y = 1 \cdot y'$: Similar to the
   previous case, but nothing gets reversed.
 \item case $x = 0 \cdot x'$ and $y = 1 \cdot y'$: In this case, we
   have $x = 0111\ldots$ and $y = 1000\ldots$.  We need to find an
   edge $0 \cdot f(\negate(111\ldots)) \to 1 \cdot f(000\ldots)$,
   which simplifies to
   $0 \cdot f(000\ldots) \to 1 \cdot f(000\ldots)$.  This edge is
   directly given in~\eqref{eq:twisted-edges-from-nodes}.
 \item case $x = 1 \cdot x'$ and $y = 0 \cdot y'$: Contradicts with
   the assumption $x+1 = y$.
 \end{itemize}
 This shows that there is a Hamiltonian path, and it is given by
 $\hat f_n$.  The definition of $f$ as in
 (\ref{eq:def-f-0},\ref{eq:def-f-1}) also shows that $f_{n+1}$
 consists of two copies of $f_n$, implying the last claim of
 \cref{thm:hamiltonian}.  In order to prove \cref{thm:total-order}, we
 need to construct $\hat g_n$.  It is enough to show that, for an edge
 from $u$ to $v$ in $\twcube n$, we have $g(u) \leqslant g(v)$.  This
 follows by straightforward induction, going through the edges in
 \cref{def:tw-graph-iter}.  But \cref{thm:total-order} implies that
 there is at most one Hamiltonian path.
\end{proof}

\begin{remark}
  Note that every vertex $v$ in $\twcube n$ is an endpoint of $n$
  non-trivial edges.  The number of zeros in the binary representation
  in the ``order number'' of $v$ (i.e.\ the value $g_n(v)$ in the
  proof of \cref{thm:total-order}) equals the number of
  \emph{outgoing} edges.  \Cref{fig:rainbows} shows this.
\end{remark}

Analogously to \cref{def:ord-graphs-def}, we can now define the
category of twisted graph morphisms:

\begin{definition}[category $\twcubecat$]~\label{def:tw-graphs-def}
  The category $\twcubecat$ has natural numbers as objects, and
  morphisms from $m$ to $n$ are graph morphisms between twisted cubes:
 \begin{align}
  & \obj{\twcubecat} \defeq \N
  & \mor{\twcubecat}(m,n) \defeq \graphhom{\twcube m}{\twcube n}
 \end{align}
\end{definition}

It is easy to see that the category $\twcubecat$ has a version of
connections.  Since we are looking for a ``twisted analogue'' of
$\shcubeop$, we need to refine it further.  In
\cref{sec:standardcubes}, we have discussed the restriction to (meet
and join)-preserving morphisms, and to dimension-preserving morphisms.
It follows directly from \cref{thm:total-order} that every morphism in
$\twcubecat$ preserves all binary meets and joins, so this condition
becomes trivial; it does not avoid connections.  However, preserving
dimensions is still a non-trivial condition which does avoid
connections.  The definition of equation~\eqref{eq:dim-preserving}
still works.

\begin{definition}[category $\twgraphdim$]
  The category $\twgraphdim$ has dimension-preserving maps between
  twisted cubes as morphisms:
 \begin{align}
  & \mathsf{obj}(\twgraphdim) \defeq \N
  & \mor{\twgraphdim}(m,n) \defeq \Sigma (g : \graphhom{\twcube m}{\twcube n}). \dimpres(g)
 \end{align}
\end{definition}

Note that the explanation of \cref{rm:dim-inj-derivable} holds for the
twisted cube category as well.

A consequence of \cref{thm:total-order} is that morphisms in
$\twgraphdim$ cannot ``swap dimensions''.  But an even stronger result
holds, namely that surjective morphisms are unique:
\begin{theorem}~\label{lem:surj-dim-unique}
 There is exactly one surjective morphism in $\twgraphdim(m,n)$ 
 for $m \geqslant n$. \\ (Clearly, there is none if $m < n$.)
\end{theorem}
\begin{proof}
  The key to the proof is \cref{thm:hamiltonian}.  Clearly, the
  Hamiltonian path in $\twcube m$ goes through all vertices.  Due to
  surjectivity, its image has to go through all vertices of
  $\twcube n$.  In other words, the $\twcube m$-Hamiltonian path has
  to be mapped to the $\twcube n$-Hamiltonian path.  Since the graph
  morphisms that we consider preserve the dimension, the only edge in
  the $\twcube m$-path which can be mapped to the single edge in the
  first dimension in the $\twcube n$-path is just this single edge in
  the first dimension in the $\twcube m$-path; i.e.\ the middle edge
  has to be mapped to the middle edge.  From here, it follows by
  induction that there can only be at most one surjective graph
  morphism.
 
  What is left to show is that there actually is a surjective graph
  morphism if $m \geqslant n$.  It is enough to construct a surjective
  graph morphism $f : \twgraphdim(n+1,n)$, from where we get any other
  by $(m-n)$-fold composition ($0$-fold composition is the identity).
  Such a graph morphism is given by
 \begin{align}
  f(x_0 \ldots x_{n-1} x_n) \defeq (x_0 \ldots x_{n-1}).
 \end{align}
 Since the directions of the edges do not depend on the very last
 dimension, this works (cf.~\cref{def:tw-cube-graph-nonrec}).
\end{proof}

An important consequence of the above result is that there is a unique
way to degenerate a twisted cube.  We do not go into the details here,
but see the conclusions at the end of the paper.  For now, we go into
a different direction.

Let us write $\intv$ (``interval'') for the finite set
$\{0,1,\star\}$.  Of course, $\intv$ is isomorphic to $\finset 3$, but
referring to the last element as $\star$ helps the intuition, we hope.

\begin{definition}
  A \emph{face} of the twisted $n$-cube $\twcube n$ is a function
  $f : \finset n \to \intv$.  The \emph{dimension} of a face, written
  $\dim(f)$, equals the number of times $f$ takes $\star$ as value
  (i.e.\ the size of $f^{-1}(\star)$).  The type of faces of dimension
  $k$ is written as $\mathsf{faces}(n, k)$.
\end{definition}

The face $f : \finset n \to \intv$ represents the full subgraph of
$\twcube n$ of vertices on which $f$ ``matches'' (a vertex
$x_0x_1\ldots{}x_{n-1}$ is matched if, for every $i$, we have
$f(i) = x_i$ or $f(i) = \star$).

\begin{lemma}~\label{lem:image-is-face}
 The image of $f : \mor{\twgraphdim}(m,n)$ is a face.
\end{lemma}
\begin{proof}
  This follows from the property of preserving the dimension as
  defined in~\eqref{eq:dim-preserving}.
\end{proof}

\begin{lemma}~\label{lem:faces-only-inj}
 The $m$-faces are the only injective maps $\twgraphdim(m,n)$:
 \begin{equation}
  \mathsf{faces}(n, m) \, \simeq \, \Sigma(f : \twgraphdim(m,n)).\isinj(f).
 \end{equation}
\end{lemma}
\begin{proof}
  Every face gives rise to a canonical injective dimension-preserving
  morphism in the sense of \cref{def:ord-dim-pres}, as dictated by the
  inclusion of the full subgraph that the face represents into
  $\twcube n$.  The fact that these are the only ones follows from
  \cref{thm:total-order} (we cannot ``swap dimensions'') and
  \cref{lem:image-is-face}.
\end{proof}

As with~\cref{thm:total-order} before, \cref{lem:faces-only-inj} is a
result which is usually found in simplicial structures, but not in
cubical ones.  In any case, we now easily get:
\begin{lemma}[factorisation of dimension preserving
  morphisms]~\label{lem:factoring-dim-pres} Given a morphism
  $f : \twgraphdim(m,n)$, there is exactly one way to write it as the
  composition $f = \mathsf{inj}(f) \circ \mathsf{surj}(f)$ of a
  surjective dimension preserving graph morphism followed by an
  injective one.  This means that the map
 \begin{align}
  & \hspace*{-.3cm} \big(\Sigma (k : \N)  . \left(\Sigma (h : \twgraphdim(k,n)). \isinj(h)\right) \times \left(\Sigma (g : \twgraphdim(m,k)). \issurj(g)\right) \big) \to \twgraphdim(m,n)
\label{eq:factorisation}\\
  & \hspace*{-.3cm} (k,(h,i),(g,s)) \mapsto h \circ g
 \end{align}
 is an equivalence.  Moreover, morphisms $\twgraphdim(m,n)$ are in
 1-to-1 correspondence with faces of $\twcube n$ of dimension
 $\leqslant m$.
\end{lemma}
\begin{proof}
  A consequence of \cref{lem:image-is-face} is that the factorisation
  on the level of sets of vertices works.  The second claim follows
  from the first: In~\eqref{eq:factorisation}, the $k$ and the
  surjective map are uniquely determined (i.e.\ contractible
  components) by \cref{lem:surj-dim-unique}.  By
  \cref{lem:faces-only-inj}, injective maps correspond to faces.
\end{proof}

\begin{remark}
  It follows from \cref{lem:factoring-dim-pres} and the proof of
  \cref{lem:surj-dim-unique} that all the non-empty fibres of a
  dimension-preserving morphism between twisted cubes have the same
  size. The reverse is the case as well: a morphism between twisted
  graphs where all non-empty fibres have the same size is
  dimension-preserving.
\end{remark}

Another consequence of the above results is that $\twgraphdim$ can be
given the structure of a \emph{Reedy category}
(cf.~\cite{hirschhorn2009model}).  Recall that a Reedy category is a
category $R$ with a degree function $d : \obj{\twgraphdim} \to \N$ and
two subcategories $R^+$ and $R^-$, such that:\footnote{Degrees can
  more generally be arbitrary ordinals, but $\N$ is sufficient in our
  case.}
\begin{itemize}
 \item both subcategories are \emph{wide}, i.e.\ contain all the objects of $R$;
 \item every nonidentity morphism in $R^+$ raises the degree;
 \item every nonidentity morphism in $R^-$ lowers the degree;
 \item and every morphism of $R$ can be written as a morphisms in
   $R^-$ followed by a morphism in $R^+$ in a unique way.
\end{itemize}
The reason why Reedy categories are interesting is that they enable
certain inductive constructions.  In the setting of type theory, they
have been discussed by Shulman~\cite{shulman6248univalence}.

\begin{theorem}
  The category $\twgraphdim$ is a Reedy category where the degree of
  an object is the object itself (recall that objects are natural
  numbers).  $\twgraphdim^+$ is the subcategory of injective
  morphisms, and $\twgraphdim^-$ is the subcategory of surjective
  morphisms.
\end{theorem}
\begin{proof}
  The first three properties are clear, and the factorisation is given
  by~\cref{lem:factoring-dim-pres}.
\end{proof}

Finally, let us record an alternative representation of the category
$\twgraphdim$ which does not go via graph morphisms.

\begin{definition}[ternary notation: category $\twternary$]
  The category $\twternary$ has natural numbers as objects, and a
  morphism from $m$ to $n$ is a function $\finset n \to \intv$ which
  takes $\star$ at most $m$ times as image:
 \begin{align}
  & \obj{\twternary} \defeq \N 
  & \mor{\twternary}(m,n) \defeq \Sigma (f : \finset n \to \intv). f^{-1}(\star) \leqslant m
 \end{align}
 The identity morphisms are the functions that are constantly $\star$.
 To define the composition of $f : \twternary(k,m)$ and
 $g : \twternary(m,n)$, we need to define a function
 $g \circ f : \finset n \to \intv$ (which is $\star$ at most $k$
 times).  We define $(g \circ f)(i)$ by recursion on $i$,
 simultaneously with the values $i'$ and $b_i$, as follows:
 \begin{align}
   (g \circ f)(i) \defeq
  \begin{cases}
    g(i)                     & \text{if } g(i) \in \{0,1\} \\
    (f(i')) \; \mathsf{xor} \; b_i & \text{if } g(i) = \star   \text{ and } f(i') \in \{0,1\} \\
    \star & \text{if } g(i) = \star  \text{ and } f(i') = \star
  \end{cases}
 \end{align}
 where \begin{itemize}
 \item $i'$ is the number of occurrences of $\star$ in the sequence
   $g(0), g(1), \ldots, g(i-1)$;
 \item $b_i$ is $1$ if the number of zeros in the sequence
   $(g \circ f)(0), (g \circ f)(1), \ldots, (g \circ f)(i-1)$ is odd,
   and $0$ if it is even.
       \end{itemize}
\end{definition}

Note that a morphism in $\mor{\twternary}(m,n)$ can be represented as
a sequence such as $01\slimstar0\slimstar10$ of length $n$ which
contains the symbol $\star$ at most $m$ times, which is why we
refer to it as \emph{ternary notation}.

\begin{remark}~\label{rm:semi-twisted-cubes} There is a category of
  twisted semi-cubes, denoted by $\twternary^+$, which is exactly the
  same as $\twternary$ except that the number of $\star$ in the
  sequence must be exactly $m$, i.e.\ ``$\leqslant$'' is changed to
  ``$=$'' in the definition of $\mor{\twternary}(m,n)$. This category
  is equivalent to the sub-category of $\twgraphdim$, denoted as
  $\twgraphdim^+$, which consists of \emph{injective}
  dimension-preserving graph homomorphisms.  Note that this
  injectivity condition is equivalent to removing the reflexive edges
  from \cref{def:tw-cube-graph-rec}.

  If we remove the expression ($\mathsf{xor} \, b_i$) in the
  definition of morphisms of $\twternary^+$, then the category becomes
  equivalent to the category of standard cubes but without
  degeneracies and swapping dimensions. In other words, the expression
  ($\mathsf{xor} \, b_i$) characterises ``twisted-ness''.
\end{remark}

\begin{theorem}
  The categories $\twgraphdim$, and $\twternary$ are isomorphic, with
  the object part being the identity.  In particular, we have:
 \begin{equation} \label{eq:tw-cubes-cat-iso}
  \mor{\twgraphdim}(m,n) \simeq \mor{\twternary}(m,n)
 \end{equation}
\end{theorem}
\begin{proof}
  As the following chain of equivalences:
  \begin{alignat*}{3}
  &&& \mor{\twgraphdim}(m,n) \\
  [\Cref{lem:factoring-dim-pres}] \quad & \simeq & \quad & \Sigma (k : \N)  . \left(\Sigma (h : \twgraphdim(k,n)). \isinj(h)\right) \times \left(\Sigma (g : \twgraphdim(m,k)). \issurj(g)\right) \\
  [\Cref{lem:surj-dim-unique}] \quad & \simeq & \quad & \Sigma (k : \N)  . \left(\Sigma (h : \twgraphdim(k,n)). \isinj(h)\right) \times (k \leqslant m) \\
  [\Cref{lem:faces-only-inj}] \quad & \simeq & \quad & \Sigma (k : \N)  . \; \mathsf{faces}(n, k) \times (k \leqslant m) \\
  [simplification] \quad & \simeq & \quad & \Sigma (f : \finset n \to \intv). f^{-1}(\star) \leqslant m \\
   \quad & \equiv & \quad & \mor{\twternary}(m,n)
 \end{alignat*}
 When transported along this isomorphism, the composition of
 $\twgraphdim$ gets mapped to the composition of $\twternary$, as
 required.
\end{proof}

\section{Conclusions and Future Directions}

We have suggested new representations of the BCH cube category and
introduced a category of twisted cubes.  It is natural to further
study the similarities and differences between standard and twisted
cube categories, and some new results will be presented in the
upcoming PhD thesis of the first author.

As future work, we plan to examine algebraic descriptions via
generators and relations.  Such presentations exist for many different
cube categories in the literature but, as far as we are aware, not for
the BCH cube category.  The closest suggestions available are the
presentations by Antolini~\cite{Antolini2002} and
Newstead~\cite{Clive-Newstead-cubical-sets}, which seem to be fairly
easy to adapt to the BCH cube category.  Interestingly, further
adapting the generators to the \emph{twisted} setting simplifies them
significantly, which mirrors the fact that morphisms between twisted
cubes cannot swap dimensions.  Moreover, our
\cref{lem:surj-dim-unique} implies that degeneracies are unique: there
is only one single way in which a twisted $n$-cube can be degenerated
to get a twisted $(n+1)$-cube.  A consequence is that we do not need
to impose relations between different degeneracies.

This, we hope, will make it possible to develop the higher categorical
structures that can be encoded as presheaves on the category of
twisted cubes.  An ultimate goal would be to model some form of
\emph{directed cubical type theory} mirroring the model by Bezem,
Coquand, and Huber~\cite{bezem_et_al:LIPIcs:2014:4628}.

Another possible application of our twisted cube categories might be
building a syntax for a parametric type theory or cubical type theory
without an interval as suggested by Altenkirch and
Kaposi~\cite{altenkirch_et_al:LIPIcs:2018:8473}.  A major difficulty
in their development was the presence of multiple degeneracies, a
problem which does not occur in the current work.

A further direction which may be worth exploring is to not consider
set-valued presheaves, but type-valued presheaves instead.  To
facilitate this, we can consider the category of twisted semi-cubes
mentioned in~\cref{rm:semi-twisted-cubes}.  From there, type-valued
presheaves can be encoded as Reedy-fibrant diagrams in a known
style~\cite{shulman_inversediagrams}.  We can then add a condition
reminiscent of Rezk's \emph{Segal-condition}~\cite{rezk2001model} by
stating that the projection from twisted semi-cubical types to the
sequence of types along the Hamiltonian path is an equivalence.  This
corresponds to saying that the partial $n$-cube with missing inner
part and lid (cf.~\cref{fig:3-cube-tw}) have a contractible type of
fillers.  It seems that this could be a first step towards the
construction of composition and higher coherences, although further
conditions seem to be necessary.  The relation to the \emph{(complete)
  semi-Segal types} by Capriotti and others
~\cite{ann-cap-kra:two-level,paolo:thesis,capKra_semisegal} remains to
be studied.

\subparagraph*{Acknowledgements} We would like to thank Paolo
Capriotti and Jakob von Raumer.  Both offered many suggestions during
fruitful exchanges.  In particular, the initial observation on which
\cref{thm:total-order} is based was suggested by them, and the idea of
considering graph morphisms was found in one of our many interesting
discussions.  The first author would like to thank his PhD supervisor,
Thorsten Altenkirch, for helpful discussions.  We are also grateful
to the participants of TYPES'19 in Oslo and the summer school on
HTT/UF in Leeds. We thank in particular Emily Riehl, Christian
Sattler, and Steve Awodey for their help and their comments.  Special
thanks go to Andreas Nuyts, who has pointed out a mistake in an
earlier draft of this paper, and to the anonymous reviewers for their
careful reading and comments.

\bibliography{bibliography}

\end{document}